\documentclass[journal]{IEEEtran}
\usepackage{amsmath,amssymb,amsfonts,bm}
\usepackage{graphicx}
\usepackage{xcolor}
\usepackage{algorithm}
\usepackage{algorithmic}
\usepackage{cite}
\usepackage{etoolbox}
\newif\ifshowrevisions
\ifshowrevisions
  \colorlet{revisioncolor}{blue}
\else
  \colorlet{revisioncolor}{black}
\fi
\newcommand{\rev}[1]{{\color{revisioncolor}#1}}
\makeatletter
\let\originalruledcaption\floatc@ruled
\renewcommand{\floatc@ruled}[2]{{\color{black}\originalruledcaption{#1}{#2}}}
\makeatother

\let\originalbibitem\bibitem
\renewcommand{\bibitem}[2][]{%
  \ifstrequal{#2}{monemi2026beamsquint}{\color{revisioncolor}}{\color{black}}%
  \ifstrempty{#1}{\originalbibitem{#2}}{\originalbibitem[#1]{#2}}}

\begin{document}

\title{Wideband Hybrid Beamforming via Programmable Unitary Mixer--TTD RF Networks}
\author{Nikola Zlatanov, Damir Salakhov, Zhen Gao
\thanks{N.~Zlatanov and D.~Salakhov are with Innopolis University, Innopolis, 420500, Russia (e-mails n.zlatanov@innopolis.ru).}
\thanks{Zhen Gao is with the Advanced Research Institute of Multidisciplinary Science, Beijing Institute of Technology, Beijing 100081, China}
}

\maketitle

\begin{abstract}
This paper proposes a wideband hybrid beamforming architecture whose analog front-end consists of fixed unitary mixers interlaced with programmable diagonal phase-shift/true-time-delay (TTD) layers. Under ideal passive-component assumptions, the analog beamformer is semi-unitary on every OFDM subcarrier, so the injected RF-chain power reaches the antenna ports without analog-network loss. We design the phase shifts and delays of the programmable diagonal phase-shift/TTD layers  for OFDM sum-rate maximization. \rev{Numerical results show that increasing mixer--TTD depth improves the sum rate under ideal component assumptions and a common injected RF-power budget. With component losses included, a shallow design retains an advantage over the evaluated hybrid baselines at low-to-moderate powers, motivating deeper implementations with low accumulated loss.}
\end{abstract}

\begin{IEEEkeywords}
Wideband hybrid beamforming, OFDM, true time delay, programmable unitary RF networks, injected power.
\end{IEEEkeywords}

\section{Introduction}
Hybrid analog--digital beamforming is a central architecture for large mmWave and sub-THz arrays because fully-digital precoding requires one RF chain per antenna, which is often infeasible in cost, power consumption, and routing complexity \cite{sohrabi2016hybrid}.

   In wideband
orthogonal frequency-division multiplexing (OFDM), however, the analog stage must accommodate frequency-dependent
array responses and multiuser channel subspaces; otherwise,
frequency-flat phase-shifter networks suffer from beam squint
and degraded spatial multiplexing \cite{alkhateeb2016frequency,sohrabi2017ofdm,wang2019beamsquint}.
 \rev{Recent true-time-delay (TTD) assisted architectures address this issue
by providing frequency-dependent phase control for wideband
beamforming~\cite{dai2022delayphase,wang2024ttdconfigurations,
wang2025beamfocusing,salakhov2026ttdriemannian,monemi2026beamsquint}.}

A separate issue is the transfer of power from the RF chains through the passive analog RF network to the antenna ports. Even under ideal passive-component assumptions, an analog RF front-end can be contractive (lossy) \cite{garcia2016rf,zlatanov2026unitary}; for example, the fully connected splitter--phase-shifter--combiner network is strongly contractive under this model \cite{garcia2016rf,zlatanov2026unitary}. \rev{Under ideal passive-component assumptions, lossless narrowband alternatives include} Butler/DFT beamforming networks \cite{garcia2016rf}, the sub-connected architecture of \cite{yu2016alternating}, and programmable-unitary RF networks \cite{zlatanov2026unitary}.

This paper extends the narrowband programmable-unitary RF network of \cite{zlatanov2026unitary} to wideband OFDM. The proposed analog beamformer is a programmable passive RF network with fixed unitary mixers interlaced with programmable phase/TTD layers. It is semi-unitary on each subcarrier and therefore transfers injected RF-chain power losslessly to the antenna ports under the ideal passive-component model. \rev{The mixer--TTD depth controls the programmable degrees of freedom. We examine the resulting ideal-component performance and the rate retained when each hybrid architecture incurs its corresponding component losses.}

The contributions are: 1) a wideband mixer--TTD RF architecture that preserves injected power tone by tone under \rev{ideal passive-component assumptions}; 2) an OFDM sum-rate formulation under an average injected RF-chain-power constraint; and \rev{3) comparisons with fully connected TTD (FC-TTD), partially connected TTD (PC-TTD), hybrid-forward-and-backward TTD (HFB-TTD), and phase-shifter-only baselines, including hardware scaling and insertion loss under a common injected-power budget.}

\section{Wideband OFDM System Model and Injected-Power Constraint}
\subsection{Transmitter, channel, and received signal}
Consider a downlink OFDM transmitter with $N$ antennas and $r$ RF chains, where $r\ll N$. The transmitter serves $S$ scheduled single-antenna users, with one data stream per user and $S\le r$. The system uses $K$ active OFDM subcarriers. With carrier frequency $f_c$ and subcarrier spacing $\Delta f$, the frequency of subcarrier $k$ is
\begin{align}
    f_k &= f_c + \xi_k, &
    \xi_k &= \left(k-\frac{K+1}{2}\right)\Delta f, &
    B&=K\Delta f ,
\end{align}
where $B$ denotes the active OFDM bandwidth.  

The channel on tone $k$ is
\begin{equation}
    \mathbf H[k]=[\mathbf h_1[k],\ldots,\mathbf h_S[k]]\in\mathbb C^{N\times S},
\end{equation}
where $\mathbf h_s[k]\in\mathbb C^{N}$ is the channel from the transmit array to user $s$. The hybrid precoder on tone $k$ is
$
    \mathbf F[k] = \mathbf F_{\mathrm{RF}}[k]\mathbf F_{\mathrm{BB}}[k],
$
where $\mathbf F_{\mathrm{RF}}[k]\in\mathbb C^{N\times r}$ is the analog RF-domain beamformer and $\mathbf F_{\mathrm{BB}}[k]\in\mathbb C^{r\times S}$ is the digital baseband precoder. 

Let $\mathbf s[k]\in\mathbb C^{S}$ denote the data vector on tone $k$, with
$\mathbb E[\mathbf s[k]\mathbf s[k]^H]=\mathbf I_S$. The RF-chain output, antenna excitation, and received vector are
\begin{align}
    \mathbf x[k] &= \mathbf F_{\mathrm{BB}}[k]\mathbf s[k],\nonumber\\
    \mathbf x_{\mathrm{ant}}[k] &= \mathbf F_{\mathrm{RF}}[k]\mathbf F_{\mathrm{BB}}[k]\mathbf s[k],\nonumber\\
    \mathbf y[k] &= \mathbf H[k]^H\mathbf F_{\mathrm{RF}}[k]\mathbf F_{\mathrm{BB}}[k]\mathbf s[k]+\mathbf n[k],
\end{align}
where $\mathbf n[k]\sim\mathcal{CN}(\mathbf 0,\sigma^2\mathbf I_S)$ and $\sigma^2$ is the post-FFT receiver-noise variance per active subcarrier at each single-antenna user.  

\subsection{Injected and antenna-excitation power}
The power constraint is imposed on the average RF-chain power injected into the passive analog RF network. The average per-tone injected power by the RF-chains is given by
\begin{equation}
    \label{eq:pinj}
    P_{\mathrm{inj}}
    =
    \frac{1}{K}\sum_{k=1}^{K}
    \mathbb E\!\left[\|\mathbf x[k]\|_2^2\right]
    =
    \frac{1}{K}\sum_{k=1}^{K}
    \|\mathbf F_{\mathrm{BB}}[k]\|_F^2 .
\end{equation}
In this paper, all considered beamforming architectures   are constrained by
$
    P_{\mathrm{inj}}\le P_T ,
$
\rev{where $P_T=P_{\mathrm{tot}}/K$ is the per-tone budget and $P_{\mathrm{tot}}$ is the total injected RF-power budget over the occupied band.}
The corresponding average antenna-excitation power is
\begin{align}
    \label{eq:pant}
    P_{\mathrm{ant}}
    &=
    \frac{1}{K}\sum_{k=1}^{K}
    \mathbb E\!\left[\|\mathbf x_{\mathrm{ant}}[k]\|_2^2\right]\nonumber\\
    &=
    \frac{1}{K}\sum_{k=1}^{K}
    \mathrm{tr}\!\left(
    \mathbf F_{\mathrm{BB}}[k]^H
    \mathbf F_{\mathrm{RF}}[k]^H
    \mathbf F_{\mathrm{RF}}[k]
    \mathbf F_{\mathrm{BB}}[k]
    \right).
\end{align}
Thus, if the analog beamformer is semi-unitary on every subcarrier,
$\mathbf F_{\mathrm{RF}}[k]^H\mathbf F_{\mathrm{RF}}[k]=\mathbf I_r$, then
$P_{\mathrm{ant}}=P_{\mathrm{inj}}$ under the ideal passive-component assumptions. By contrast, for a passive contractive analog RF network satisfying
$\mathbf F_{\mathrm{RF}}[k]^H\mathbf F_{\mathrm{RF}}[k]\preceq\mathbf I_r$, the delivered antenna-excitation power satisfies $P_{\mathrm{ant}}\le P_{\mathrm{inj}}$.

\rev{The RF chains supply power to the analog beamformer network, which then transfers this power to the antenna ports. Even an ideal analog beamformer network, such as the FC splitter--combiner network, can divert part of the power supplied by the RF chains to terminated ports, thereby leaving the antenna ports with less power than that supplied by the RF chains. As a result, comparing hybrid beamforming architectures under equal antenna-excitation power may compare them at different levels of power supplied by the RF chains, concealing the effect of power expenditure on data rate. On the other hand, comparing hybrid beamforming architectures under the same power supplied by the RF chains makes the comparison fair and exposes the effect of power expenditure on rate.}

\subsection{OFDM sum spectral efficiency}
Define $
    \mathbf F_{\mathrm{eff}}[k]\triangleq \mathbf F_{\mathrm{RF}}[k]\mathbf F_{\mathrm{BB}}[k]$ and 
   $ \mathbf G[k]\triangleq \mathbf H[k]^H\mathbf F_{\mathrm{eff}}[k].$
Each user applies single-user decoding and treats the streams intended for the other users as interference. With $g_{s,j}[k]$ denoting the $(s,j)$th entry of $\mathbf G[k]$, the SINR of user $s$ on subcarrier $k$ is $
    \mathrm{SINR}_s[k]
    =
    \frac{|g_{s,s}[k]|^2}
    {\sum_{j\neq s}|g_{s,j}[k]|^2+\sigma^2}.
$
The average OFDM sum spectral efficiency is then
\begin{equation}
    \label{eq:ofdm_sumrate_common}
    R_{\Sigma}^{\mathrm{OFDM}}
    =
    \frac{1}{K}\sum_{k=1}^K\sum_{s=1}^S
    \log_2\bigl(1+\mathrm{SINR}_s[k]\bigr),
\end{equation}
measured in bit/s/Hz over the active OFDM bandwidth.

\section{Programmable Unitary Mixer--TTD RF Network}
\subsection{Layered mixer--TTD signal model}
Let $\mathbf W\in\mathbb C^{N\times N}$ be a fixed unitary mixing matrix,
$\mathbf W^H\mathbf W=\mathbf I_N$, such as a DFT or Butler matrix. Let
$M_{\mathrm{lay}}$ denote the number of programmable diagonal phase/TTD layers, and define
$M\triangleq M_{\mathrm{lay}}$ for compact notation in products and subscripts. The full
$N\times N$ analog processor on subcarrier $k$ is modeled as
\begin{align}
    \label{eq:common_Xk}
    \mathbf X[k]
    = \mathbf W\mathbf D_M[k]\mathbf W\mathbf D_{M-1}[k]
    \cdots \mathbf W\mathbf D_1[k]\mathbf W,
\end{align}
where
\begin{align}
    \mathbf D_m[k]
    = \mathrm{diag}\bigl(d_{1,m}[k],\ldots,d_{N,m}[k]\bigr),
    \qquad m=1,\ldots,M,
\end{align}
is the $m$th diagonal phase/TTD layer with
\begin{align}
    \label{eq:dnm_wb}
    d_{n,m}[k]
    &= e^{j\phi_{n,m}}e^{-j\omega_k\tau_{n,m}},
    \qquad
    \omega_k\triangleq 2\pi(f_k-f_c)=2\pi\xi_k.
\end{align}
In \eqref{eq:dnm_wb}, $\phi_{n,m}\in[0,2\pi)$ is a \rev{programmable, frequency-independent phase shift} and
$\tau_{n,m}\in[0,\tau_{\max}]$ is a programmable TTD. Since matrix products are applied from right to left, the signal first passes through the rightmost fixed mixer, then through the first diagonal phase/TTD layer, and so on until the final fixed mixer. Thus, the architecture contains $M_{\mathrm{lay}}$ programmable diagonal layers and $M_{\mathrm{lay}}+1$ fixed unitary mixing stages.

An illustration of the proposed hybrid beamforming architecture can be obtained from
Fig.~1 in \cite{zlatanov2026unitary} by replacing each  phase-shifter element by a cascade of a programmable phase shifter and a programmable TTD element.

In the proposed hybrid architecture, only $r$ input ports of the $N$-port processor are driven by RF chains; the remaining ports are matched. Let
$\mathbf E_r=[\mathbf e_1,\ldots,\mathbf e_r]\in\mathbb R^{N\times r}$ be the selection matrix for the driven ports. Then the proposed analog beamformer on subcarrier $k$ is
\begin{equation}
    \label{eq:common_frf}
    \mathbf F_{\mathrm{RF}}[k]=\mathbf X[k]\mathbf E_r .
\end{equation}

\rev{The proposed analog beamformer uses $NM_{\mathrm{lay}}$ programmable phase shifters and $NM_{\mathrm{lay}}$ programmable TTD elements, giving $2NM_{\mathrm{lay}}$ real programmable parameters.} Frequency selectivity arises because the same delay $\tau_{n,m}$ yields the tone-dependent phase $-\omega_k\tau_{n,m}$.

\subsection{Subcarrier-wise semi-unitarity}
In the proposed analog beamformer, each fixed mixing stage is unitary and each diagonal phase/TTD layer has unit-modulus entries. Hence,
\begin{equation}
    \label{eq:X_unitary}
    \mathbf X[k]^H\mathbf X[k]=\mathbf I_N,
    \qquad k=1,\ldots,K .
\end{equation}
Using \eqref{eq:common_frf}, the analog beamformer therefore satisfies
\begin{align}
    \label{eq:frf_semiunitary}
    \mathbf F_{\mathrm{RF}}[k]^H\mathbf F_{\mathrm{RF}}[k]
    &=
    \mathbf E_r^H\mathbf X[k]^H\mathbf X[k]\mathbf E_r\nonumber\\
    &=
    \mathbf E_r^H\mathbf E_r
    =
    \mathbf I_r,
    \qquad k=1,\ldots,K .
\end{align}
Thus, the proposed RF network produces a semi-unitary $N\times r$ analog beamformer on every subcarrier. By \eqref{eq:pant}, this implies that the antenna-excitation power equals the injected RF-chain power under the ideal matched, uncoupled antenna model.  

\noindent\rev{\textit{Remark:} For the proposed hybrid beamforming architecture at a fixed depth $M$, the maximum sum rate under ideal passive-component assumptions provides an upper bound on the corresponding performance with the passive-component losses considered, cf. Section~\ref{sec:numerical_evaluation}.}

\section{Hybrid Precoder Design}
\subsection{Problem formulation}
The analog design variables are the phase and delay collections
$\Phi\triangleq\{\phi_{n,m}\}$ and
$\mathcal T\triangleq\{\tau_{n,m}\}$, while the digital variables are the
tone-dependent baseband precoders
$\{\mathbf F_{\mathrm{BB}}[k]\}_{k=1}^K$. For compactness, write
$\mathbf X[k;\Phi,\mathcal T]$ for the mixer--TTD processor generated by
\eqref{eq:common_Xk}--\eqref{eq:dnm_wb}. \rev{The injected-power hybrid design problem, with constraints imposed for all $k,n,m$, is}
\begin{align}
    \label{eq:mod1_problem}
    \max_{\Phi,\mathcal T,\{\mathbf F_{\mathrm{BB}}[k]\}}
    \quad & R_{\Sigma}^{\mathrm{OFDM}}\\
    \mathrm{s.t.}\quad
    &\mathbf F_{\mathrm{RF}}[k]=\mathbf X[k;\Phi,\mathcal T]\mathbf E_r,\nonumber\\
    &\frac{1}{K}\sum_{k=1}^K\|\mathbf F_{\mathrm{BB}}[k]\|_F^2\le P_T,\nonumber\\
    &0\le\tau_{n,m}\le\tau_{\max},\qquad
    \phi_{n,m}\in[0,2\pi).\nonumber
\end{align}
\rev{The semi-unitarity condition follows from the analog beamformer construction for every feasible phase and delay assignment.}
This problem is nonconvex.  We
therefore seek high-quality locally optimized designs using a warm-started
alternating procedure.  

\subsection{\rev{Initialization from a fully digital precoder}}
Let $\{\mathbf F_{\mathrm{FD}}[k]\}_{k=1}^K$, with
$\mathbf F_{\mathrm{FD}}[k]\in\mathbb C^{N\times S}$, denote the fully digital
precoders obtained by \rev{WMMSE} without analog constraints under the
same average power budget
$\frac{1}{K}\sum_{k=1}^{K}\|\mathbf F_{\mathrm{FD}}[k]\|_F^2\le P_T$.
The analog beamformer $\mathbf F_{\mathrm{RF}}[k]$, and thereby the phase and
delay variables, are initialized by maximizing
$\rev{J_{\mathrm{init}}}(\Phi,\mathcal T)$, given by
\begin{align}
    \label{eq:Jinit}
    \rev{J_{\mathrm{init}}}(\Phi,\mathcal T)
    &\triangleq
    \frac{1}{K E_{\mathrm{FD}}}\sum_{k=1}^K
    \left\|\mathbf F_{\mathrm{RF}}[k]^H\mathbf F_{\mathrm{FD}}[k]\right\|_F^2,
\end{align}
where  $E_{\mathrm{FD}}
     \triangleq
    \frac{1}{K}\sum_{k=1}^K\|\mathbf F_{\mathrm{FD}}[k]\|_F^2 $.
Since $\mathbf F_{\mathrm{RF}}[k]$ is semi-unitary,
$\|\mathbf F_{\mathrm{RF}}[k]^H\mathbf F_{\mathrm{FD}}[k]\|_F^2\le
\|\mathbf F_{\mathrm{FD}}[k]\|_F^2$, so $\rev{J_{\mathrm{init}}}\in[0,1]$.
Thus $\rev{J_{\mathrm{init}}}$ is the fraction of fully digital precoder energy captured by the mixer--TTD analog subspace. This stage only initializes the variables; the final design maximizes \eqref{eq:ofdm_sumrate_common}. \rev{We first optimize the phases with fixed delays, then jointly optimize phases and delays, retaining the largest $J_{\mathrm{init}}$.}

\subsection{Fixed-analog WMMSE digital update}
For fixed analog matrices, $\{\mathbf F_{\mathrm{RF}}[k]\}_{k=1}^K$, the digital precoder matrices, $\{\mathbf F_{\mathrm{BB}}[k]\}_{k=1}^K$, are updated by the standard
WMMSE block \cite{shi2011iteratively} applied to the effective RF-chain-domain
channels \rev{$\widetilde{\mathbf h}_s[k]\triangleq\mathbf F_{\mathrm{RF}}[k]^H\mathbf h_s[k]\in\mathbb C^{r}$.}
Let $\mathbf f_s[k]$ denote the $s$th column of
$\mathbf F_{\mathrm{BB}}[k]$. With the scalar receiver convention
$\widehat{s}_s[k]=u_s[k]^*y_s[k]$, the MMSE receiver, MSE, and WMMSE weight
are
{\color{revisioncolor}
\begin{align}
    \eta_s[k]&=\sigma^2+\sum_j|g_{s,j}[k]|^2,
    &u_s[k]&=\frac{g_{s,s}[k]}{\eta_s[k]},\nonumber\\
    e_s[k]&=1-\frac{|g_{s,s}[k]|^2}{\eta_s[k]},
    &\alpha_s[k]&=e_s[k]^{-1}.
    \label{eq:wmmse_receiver_weight}
\end{align}}
For fixed $\{u_s[k],\alpha_s[k]\}$, define
\begin{align}
    \mathbf A[k]
    &=
    \sum_{s=1}^{S}
    \alpha_s[k]|u_s[k]|^2
    \widetilde{\mathbf h}_s[k]\widetilde{\mathbf h}_s[k]^H,
    \nonumber\\
    \mathbf B[k]
    &=
    \bigl[
    \alpha_1[k]u_1[k]\widetilde{\mathbf h}_1[k],
    \ldots,
    \alpha_S[k]u_S[k]\widetilde{\mathbf h}_S[k]
    \bigr].
    \label{eq:wmmse_AB}
\end{align}
The digital precoder update is then
\begin{equation}
    \label{eq:wmmse_digital_update}
    \mathbf F_{\mathrm{BB}}[k]
    =
    \left(\mathbf A[k]+\lambda\mathbf I_r\right)^{-1}\mathbf B[k],
    \qquad k=1,\ldots,K,
\end{equation}
where the same Lagrange multiplier $\lambda\ge0$ is used for all subcarriers. \rev{At $\lambda=0$, the inverse is interpreted as the Moore--Penrose pseudoinverse, giving the minimum-norm solution.}
If the unconstrained update with $\lambda=0$ satisfies the average injected
power constraint, then $\lambda=0$ is used. Otherwise, $\lambda$ is chosen by
scalar bisection so that \rev{$\frac{1}{K}\sum_{k=1}^{K}\|\mathbf F_{\mathrm{BB}}[k]\|_F^2=P_T$.}
The receiver, weight, and precoder updates are iterated for the prescribed
number of WMMSE inner iterations. \rev{Each fresh WMMSE solve starts from the maximum-ratio or regularized zero-forcing precoder giving the higher rate after injected-power normalization. Digital updates are retained only if the rate does not decrease.}

\subsection{Projected analog-gradient update}
\rev{For fixed digital precoders, we update the phase and delay variables by projected first-order ascent.}

{\color{revisioncolor}
Write $R=R_{\Sigma}^{\mathrm{OFDM}}$ and
$\iota_s[k]=\eta_s[k]-|g_{s,s}[k]|^2$. The RF-matrix gradients of $R$ and $J_{\mathrm{init}}$ are
\begin{align}
 [\mathbf Z[k]]_{s,j}
 &=g_{s,j}[k]\left(\frac{1}{\eta_s[k]}-
 \frac{\mathbf 1_{\{j\ne s\}}}{\iota_s[k]}\right),\nonumber\\
 \boldsymbol\Gamma_R[k]
 &=\frac{2}{K\ln 2}\mathbf H[k]\mathbf Z[k]\mathbf F_{\mathrm{BB}}[k]^H,
 \label{eq:rf_rate_gradient}\\
 \boldsymbol\Gamma_{\mathrm{init}}[k]
 &=\frac{2}{KE_{\mathrm{FD}}}\mathbf F_{\mathrm{FD}}[k]
 \mathbf F_{\mathrm{FD}}[k]^H\mathbf F_{\mathrm{RF}}[k].
 \label{eq:rf_init_gradient}
\end{align}
For either objective $\mathcal J\in\{R,J_{\mathrm{init}}\}$, the convention is
$\mathrm d\mathcal J=\sum_k\operatorname{Re}\operatorname{tr}
(\boldsymbol\Gamma_{\mathcal J}[k]^H\mathrm d\mathbf F_{\mathrm{RF}}[k])$.
Factor $\mathbf F_{\mathrm{RF}}[k]=\mathbf L_m[k]\mathbf D_m[k]\mathbf R_m[k]$,
where $\mathbf L_m[k]$ and $\mathbf R_m[k]$ contain the factors to the left and right of layer $m$, including $\mathbf E_r$ in the latter. Define
$\mathbf C_m[k]=\mathbf L_m[k]^H\boldsymbol\Gamma_{\mathcal J}[k]\mathbf R_m[k]^H$
and $q_{n,m}[k]=\operatorname{Im}\{d_{n,m}[k]^*[\mathbf C_m[k]]_{n,n}\}$.
Then
\begin{align}
 \frac{\partial\mathcal J}{\partial\phi_{n,m}}
 &=\sum_k q_{n,m}[k],&
 \frac{\partial\mathcal J}{\partial\tau_{n,m}}
 &=-\sum_k\omega_k q_{n,m}[k].
 \label{eq:analog_explicit_gradients}
\end{align}
Only the diagonal of $\mathbf C_m[k]$ is needed.
We use dimensionless delay coordinates
\begin{equation}
 \Psi=\pi B\mathcal T,\qquad
 \nabla_{\Psi}\mathcal J=(\pi B)^{-1}\nabla_{\mathcal T}\mathcal J.
 \label{eq:scaled_delay}
\end{equation}
For each active variable collection $x\in\{\Phi,\Psi\}$, let
\begin{equation}
 \mathbf g_x=\frac{\nabla_x\mathcal J}
 {\max\{1,\|\nabla\mathcal J\|_2/G_{\max}\}},
 \label{eq:gradient_clipping}
\end{equation}
where $\nabla\mathcal J$ stacks the active gradients and $G_{\max}>0$ is the clipping threshold. The adaptive direction is
\begin{equation}
\begin{gathered}
 \mathbf m_x\leftarrow\beta_1\mathbf m_x+(1-\beta_1)\mathbf g_x,\\
 \mathbf v_x\leftarrow\beta_2\mathbf v_x+(1-\beta_2)\mathbf g_x^{\odot 2},\\
 \widehat{\mathbf m}_x=\frac{\mathbf m_x}{1-\beta_1^{t_x}},\quad
 \widehat{\mathbf v}_x=\frac{\mathbf v_x}{1-\beta_2^{t_x}},\quad
 \mathbf p_x=\frac{\widehat{\mathbf m}_x}{\sqrt{\widehat{\mathbf v}_x}+\epsilon}.
\end{gathered}
 \label{eq:adaptive_direction}
\end{equation}
Operations are elementwise, $\beta_1,\beta_2\in[0,1)$, $\epsilon>0$, and $t_x$ counts updates of $x$ since its moments were zeroed. With separate nominal steps $\mu_\Phi,\mu_\Psi$, a trial update is
\begin{align}
 \Phi^+&=\operatorname{wrap}_{2\pi}
 (\Phi+\alpha\mu_\Phi\mathbf p_\Phi),\nonumber\\
 \Psi^+&=\operatorname{clip}_{[0,\pi B\tau_{\max}]}
 (\Psi+\alpha\mu_\Psi\mathbf p_\Psi),
 \label{eq:analog_projected_update}
\end{align}
where wrapping maps phases into $[0,2\pi)$ and clipping truncates each scaled delay to its feasible interval. Initialization uses $\mathcal J=J_{\mathrm{init}}$ and $\alpha=1$, omitting the $\Psi$ update and its moments during the phase-only stage.

With digital precoders fixed, set $\mathcal J=R$,
$\Delta_\Phi=\alpha\mu_\Phi\mathbf p_\Phi$, $\Delta_\Psi=\Psi^+-\Psi$,
and $\rho=\sum_x\langle\mathbf g_x,\Delta_x\rangle$, using the real inner product.
Accept a trial if
\begin{equation}
 \rho>0,\qquad R^+\ge R+c_{\mathrm A}\rho,\qquad 0<c_{\mathrm A}<1.
 \label{eq:rate_acceptance}
\end{equation}
Test $\alpha=1,1/2,\ldots$ until \eqref{eq:rate_acceptance} holds or the trial limit is reached.
If unsuccessful, restart the search with
$\mathbf p_x=\mathbf g_x/(\sqrt{\widehat{\mathbf v}_x}+\epsilon)$;
retain the current variables if both searches fail.
Check
\begin{equation}
 \frac{R^{(i)}-R^{(i-L_{\mathrm{chk}})}}{\max\{R^{(i-L_{\mathrm{chk}})},\epsilon_0\}}
 \le\varepsilon_{\mathrm{stop}}
 \label{eq:rate_stopping}
\end{equation}
every $L_{\mathrm{chk}}$ iterations with $i\ge I_{\min}$, and stop when it holds or $I_{\max}$ is reached.
Here $R^{(i)}$ is the rate after digital refinement and $\epsilon_0>0$.
}

\noindent\emph{Algorithmic Form:} The entire algorithm for solving \eqref{eq:mod1_problem} is given by Algorithm~\ref{alg:main}.

\begin{algorithm}[t]
\caption{Programming of the proposed hybrid beamformer}
\label{alg:main}
{\color{revisioncolor}
\begin{algorithmic}[1]
\STATE \textbf{Input:} $\{\mathbf H[k],\omega_k\}_{k=1}^{K}$, $\sigma^2$, $\mathbf W$, $r$, $B$, $P_T$, $M$, $\tau_{\max}$, and initialization, WMMSE, update, and stopping settings.
\STATE \textbf{Output:} $(\Phi,\mathcal T)$ and $\{\mathbf F_{\mathrm{BB}}[k]\}_{k=1}^{K}$.
\STATE Compute $\{\mathbf F_{\mathrm{FD}}[k]\}$ by \eqref{eq:wmmse_receiver_weight}--\eqref{eq:wmmse_digital_update} with $\mathbf F_{\mathrm{RF}}=\mathbf I_N$ and RF dimension $N$.
\FOR{each initialization}
    \STATE Draw random phases; set $\tau_{n,m}=\tau_{\max}/2$ and $\Psi=\pi B\mathcal T$. Zero the moments and counters.
    \STATE Optimize \eqref{eq:Jinit} using \eqref{eq:rf_init_gradient}--\eqref{eq:analog_projected_update} with $\alpha=1$: first phases only, then phases and delays jointly. Retain the largest-$J_{\mathrm{init}}$ iterate.
    \STATE Form $\mathbf F_{\mathrm{RF}}[k]$ by \eqref{eq:common_Xk}--\eqref{eq:common_frf}; initialize the digital precoders by \eqref{eq:wmmse_receiver_weight}--\eqref{eq:wmmse_digital_update}.
    \FOR{$i=1,\ldots,I_{\max}$}
        \STATE At $i=1,1+L_{\mathrm{chk}},\ldots$, zero the rate-update moments and counters.
        \STATE Compute the rate gradients by \eqref{eq:rf_rate_gradient}, \eqref{eq:analog_explicit_gradients}, and \eqref{eq:scaled_delay}; apply \eqref{eq:gradient_clipping}--\eqref{eq:rate_acceptance}, including the fallback if needed.
        \STATE Refine the digital precoders by \eqref{eq:wmmse_receiver_weight}--\eqref{eq:wmmse_digital_update}; retain nondecreasing rates and perform any scheduled additional digital refinement.
        \STATE If $i\ge I_{\min}$ and $i$ is a multiple of $L_{\mathrm{chk}}$, check \eqref{eq:rate_stopping} and stop if satisfied.
    \ENDFOR
    \STATE Apply final digital WMMSE refinement; store the joint analog/digital design and its rate \eqref{eq:ofdm_sumrate_common}.
\ENDFOR
\STATE Return the stored design with the highest rate, with $\mathcal T=\Psi/(\pi B)$.
\end{algorithmic}}
\end{algorithm}

\noindent\emph{Complexity:} With FFT-implementable mixers such as DFT or Butler matrices, each analog forward/adjoint sweep costs $\mathcal O(KM_{\mathrm{lay}}rN\log N)$, or $\mathcal O(KM_{\mathrm{lay}}rN^2)$ for dense unitary mixers. Each fixed-analog WMMSE iteration adds $\mathcal O(KNSr+KSr^2+Kr^3)$ operations, up to the scalar bisection trials for the power multiplier.

\section{Numerical Evaluation}
\label{sec:numerical_evaluation}

\begin{figure*}[!t]
    \centering
    \includegraphics[width=1\textwidth]{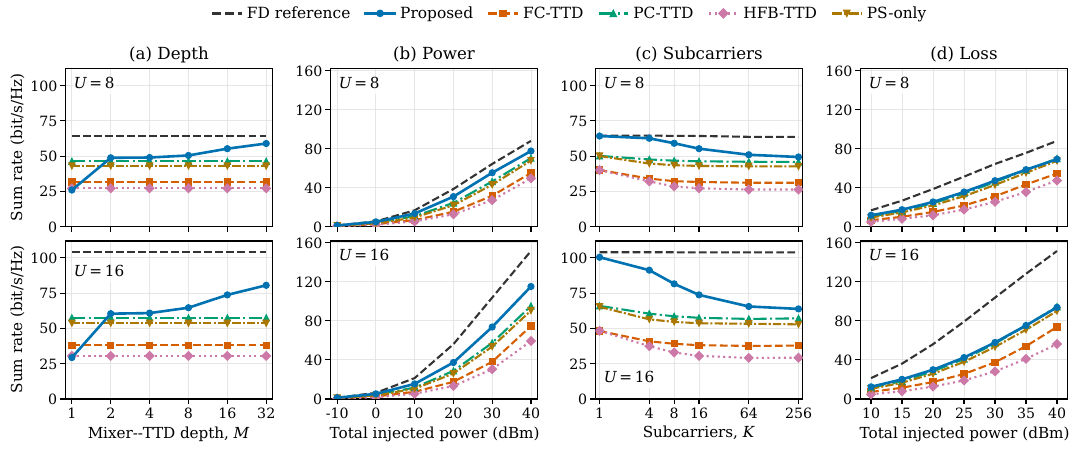}
    \caption{\rev{Average OFDM sum spectral efficiency for $U=8$ (top) and $U=16$ (bottom). Columns show (a) ideal-component depth sweep at $K=16$, $P_{\mathrm{tot}}=30$ dBm; (b) ideal-component power sweep at $M=16$, $K=16$; (c) ideal-component subcarrier sweep at $M=16$, $P_{\mathrm{tot}}=30$ dBm, with fixed $B=10$ GHz; and (d) power sweep with $M=2$, $K=16$, and component loss $\delta=0.1$ dB applied to all hybrid architectures. FD uses ideal components.}}
    \label{fig:compact_results}
\end{figure*}

\subsection{\rev{Simulation setup}}
\rev{The simulations use a half-wavelength ULA with $N=512$ antennas, $r=16$ RF chains, $f_c=100$ GHz, bandwidth $B=10$ GHz, DFT mixers, and delay limit $\tau_{\max}=2.56$ ns per physical TTD. The conventional TTD architectures use $N_T=16$ TTDs per RF chain. When varying $K$, we keep $B$ and total injected power $P_{\mathrm{tot}}$ fixed, with $\Delta f=B/K$, $P_T=P_{\mathrm{tot}}/K$, and $\sigma^2=N_0B/K$, where $N_0=10^{-20.4}$ W/Hz.}

The scheduled user load is $U\triangleq S\in\{8,16\}$, with one stream per user and equal user weights.
\rev{Channels have spherical-wave phases and frequency-dependent Friis attenuation, with one LoS path and four NLoS paths having uniform random phases and 15-dB excess attenuation. Path ranges and angles are uniform over $[5,50]$ m and $[-\pi/3,\pi/3]$, respectively, with angles measured from the array axis. Curves average five paired channel realizations per load, shared across architectures and subcarrier grids.}

\rev{Each hybrid configuration uses two initializations. Deeper proposed designs additionally use embedded shallower solutions, and fully digital precoders are refined from digital and hybrid candidates.}

\subsection{\rev{Compared architectures and component losses}}
\rev{The proposed scheme uses Algorithm~\ref{alg:main}. We compare it with fully digital (FD), FC-TTD~\cite{salakhov2026ttdriemannian}, PC-TTD~\cite{wang2025beamfocusing}, HFB-TTD~\cite{wang2024ttdconfigurations}, and subconnected PS-only~\cite{yu2016alternating} beamformers. FC-TTD and HFB-TTD are fully connected with normalization $1/\sqrt{Nr}$ for passive splitting and combining. PC-TTD and PS-only use disjoint subarrays with normalization $1/\sqrt{N/r}$. HFB-TTD uses forward cumulative delays on half the RF chains and backward cumulative delays on the remainder. All physical TTDs obey the same delay limit; serial path delays are cumulative. The PS-only phases are shared across the band.}

\rev{The (phase-shifter, TTD) counts are $(rN,rN_T)$ for FC-TTD and HFB-TTD, $(N,rN_T)$ for PC-TTD, and $(N,0)$ for PS-only.}

\rev{For the loss comparison, each complete $N$-port mixer, phase shifter, and TTD has excess loss $\delta=0.1$ dB. The path losses are $(3M+1)\delta$ for the proposed analog beamformer, $2\delta$ for FC-TTD and PC-TTD, and $\delta$ for PS-only. An HFB tap traversing $j$ TTDs incurs $(j+1)\delta$. A path loss of $L$ dB multiplies its RF amplitude by $10^{-L/20}$. Splitter/combiner excess loss is zero; their ideal power division is included in the normalizations of the baselines. We optimize the analog and digital precoders with component losses included, using the same RF-chain input-power budget for every architecture.}

\subsection{Numerical results}
\rev{Fig.~\ref{fig:compact_results}(a) shows that increasing depth improves the proposed architecture's ideal-component sum rate. At $M=8$, the proposed and FC-TTD analog beamformers have comparable total numbers of programmable elements. Column~(b) shows the ideal-component gains across injected power. At equal injected power, these comparisons capture both beamforming flexibility and power transfer through the analog beamformer.}

\rev{Column~(c) shows that the proposed architecture retains its advantage over the hybrid baselines as the fixed band is sampled more densely, with smaller rate changes between the largest subcarrier grids. The PC-TTD/PS-only comparison illustrates the benefit of delay control within the same subconnected topology.}

\rev{With component losses included, column~(d) shows an advantage for the shallow $M=2$ design at low-to-moderate powers, while PC-TTD is slightly ahead at the high-power end. PC-TTD remains close to its ideal-component performance. The higher ideal rates at greater proposed depths motivate deeper implementations with low accumulated loss.}

\section{Conclusion}
This paper proposed a wideband hybrid beamformer interlacing fixed unitary mixers with programmable phase/TTD layers. Under ideal passive-component assumptions, its analog beamformer is semi-unitary on every subcarrier, preserving injected RF-chain power. We developed an alternating algorithm for OFDM sum-rate maximization. At equal injected power, simulations show ideal-component gains over hybrid baselines and a shallow design's advantage with losses at low-to-moderate powers. These results motivate deeper implementations with low accumulated loss.

\bibliographystyle{IEEEtran}
\bibliography{wideband_paper_references}

\end{document}